\documentclass[pss]{wiley2sp} 
\usepackage{amsmath}

\begin{document}

\title{Zinc Oxide -- From Dilute Magnetic Doping to Spin Transport}

\titlerunning{ZnO -- from dilute magnetic doping to spin transport}

\author{%
  Matthias Opel\textsuperscript{\Ast,\textsf{\bfseries 1}},
  Sebastian T. B. Goennenwein\textsuperscript{\textsf{\bfseries 1}},
  Matthias Althammer\textsuperscript{\textsf{\bfseries 1}},
  Karl-Wilhelm Nielsen\textsuperscript{\textsf{\bfseries 1}},
  Eva-Maria Karrer-M\"{u}ller\textsuperscript{\textsf{\bfseries 1}},
  Sebastian Bauer\textsuperscript{\textsf{\bfseries 1}},
  Konrad Senn\textsuperscript{\textsf{\bfseries 1}},
  Christoph Schwark\textsuperscript{\textsf{\bfseries 2}},
  Christian Weier\textsuperscript{\textsf{\bfseries 2}},
  Gernot G\"{u}ntherodt\textsuperscript{\textsf{\bfseries 2,3}},
  Bernd Beschoten\textsuperscript{\textsf{\bfseries 2,3}},
  Rudolf Gross\textsuperscript{\Ast, \textsf{\bfseries 1,4}}}

\authorrunning{Matthias Opel et al.}

\mail{e-mail
  \textsf{Matthias.Opel@wmi.badw.de}, \textsf{Rudolf.Gross@wmi.badw.de}}

\institute{%
  \textsuperscript{1}\,Walther-Mei{\ss}ner-Institut,
                       Bayerische Akademie der Wissenschaften,
                       85748 Garching,
                       Germany\\
  \textsuperscript{2}\,II.~Physikalisches Institut,
                       RWTH Aachen University,
                       52056 Aachen,
                       Germany\\
  \textsuperscript{3}\,JARA-Fundamentals of Future Information Technology,
                       J\"{u}lich-Aachen Research Alliance,
                       Germany\\
  \textsuperscript{4}\,Physik-Department,
                       Technische Universit\"{a}t M\"{u}nchen,
                       85748 Garching,
                       Germany}

\received{XXXX, revised XXXX, accepted XXXX} 
\published{XXXX} 


\abstract{%
\abstcol{
  During the past years there has been renewed interest in the wide-bandgap II-VI semiconductor ZnO, triggered by promising prospects for spintronic applications. First, ferromagnetism was predicted for dilute magnetic doping. In a comprehensive investigation of ZnO:Co thin films based on the combined measurement of macroscopic and microscopic properties, we find no evidence for carrier-mediated itinerant ferromagnetism. Phase-pure, crystallographically excellent ZnO:Co is uniformly paramagnetic. Superparamagnetism arises when phase separation or defect formation occurs, due to nanometer-sized metallic precipitates. Other compounds like ZnO:(Li,Ni) and ZnO:Cu do not exhibit indication of ferromagnetism.
}{
  Second, its small spin-orbit coupling and correspondingly large spin coherence length makes ZnO suitable for transporting or manipulating spins in spintronic devices. From optical pump/optical probe experiments, we find a spin dephasing time of the order of 15\,ns at low temperatures which we attribute to electrons bound to Al donors. In all-electrical magnetotransport measurements, we successfully create and detect a spin-polarized ensemble of electrons and transport this spin information across several nanometers. We derive a spin lifetime of 2.6\,ns for these itinerant spins at low temperatures, corresponding well to results from an electrical pump/optical probe experiment.
}}

\maketitle

\section{ZnO for Spintronic Applications.}
\label{sec:intro}

Classical semiconductor-based devices rely on the controlled transport and storage of electrical charge. In semiconductor \emph{spin}tronics, however, both the electrons' charge and \emph{spin} degree of freedom are exploited, providing fascinating perspectives for novel devices with improved performance. Moreover, since the spin degree of freedom usually is coupled much more weakly to the environment than the charge degree of freedom, spin-based devices are promising for future quantum devices making use of quantum superposition and entanglement. Therefore both classical and quantum spintronics are rapidly growing areas of basic and applied research \cite{Bader2010}. More than two decades ago the exploitation of the spin degree of freedom in electronic devices started with the field of \emph{magnetoelectronics}, in which ferromagnetic metals have been used in passive spin-dependent devices like spin valves, magnetic tunnel junctions, or magnetic sensors \cite{Prinz1998}. Magnetoelectronics then has been extended to the much broader field of spintronics which is considered a replacement technology providing improved spin-active devices with the goal to encode information in single spins. The control of these spins (read, write, transport, manipulate) is realized by magnetic fields, photonic fields, electric fields, or electric currents \cite{Bader2010}.

With regard to materials and starting from metal-based devices in the late 1980s, the field of spintronics soon expanded to the use of transition-metal oxides \cite{Dietl2000,Ogale2005,Gross2006a,Bibes2007,Opel2012}. Within the past two decades, the growth of high-quality oxide thin films and heterostructures progressed enormously concerning sample and interface quality \cite{Gross2000,Bibes2011,Opel2011}. Among the various material systems for spintronic applications, the II-VI semiconductor ZnO \cite{Janotti2009} is of interest due to several reasons. First, ZnO has well been studied such that considerable knowledge on doping and defect chemistry is available \cite{Ozgur2005,Klingshirn2010}. Second, ZnO as a wide bandgap semiconductor is suitable for high power and high temperature operation \cite{Look2001}. Third, for $n$-type bulk ZnO a long room temperature spin coherence time was reported \cite{Ghosh2005}. Fourth, ZnO is able to emit light in the UV range making it attractive for a variety of optical applications \cite{Look2001}. And finally, since both hole and electron mediated ferromagnetism were reported in transition metal (TM)-substituted ZnO, it was considered to have potential for bipolar spintronics \cite{Kittilstved2006a,Nielsen2005,Chambers2006,Kittilstved2006b,Kumar2010}. However, in the past years it turned out to be unexpectedly difficult to gain control over some of the above mentioned aspects. In particular, the inability to reproducibly realize both dilute magnetic doping and $p$-type conductivity represents a huge drawback for the successful implementation of ZnO into spintronic applications \cite{Klingshirn2007} -- apart from its use as a simple transparent conducting oxide (TCO) \cite{Grundmann2010}.

In this article, we review our results concerning the ZnO system which we obtained within the priority program 1285 (``Semiconductor Spintronics'') of the German Research Foundation (DFG). In the first part (section~\ref{sec:DMS}), we focus on magnetic properties and the possibility of dilute magnetic doping with transition metal ions. Details on these aspects are published in Refs.~\cite{Opel2008,Ney2010}. In the second part (section~\ref{sec:spintronics}), we highlight spin coherence and spin dephasing properties which become important for spintronic devices with regard to the injection, transport, manipulation, and detection of spin-polarized carriers in ZnO, published in Refs.~\cite{Kuhlen2013b,Kuhlen2013c,Althammer2012}.

\section{Dilute Magnetic Doping in ZnO.}
\label{sec:DMS}

Integrating the spin degree of freedom into semiconductors via doping/substitution with magnetic ions is a key prerequisite for the development of novel functional materials, combining ferromagnetic long-range order and semiconducting behavior in a single phase. Such materials could overcome the well-known conductance mismatch problem for diffusive spin injection from a conventional ferromagnetic metallic electrode \cite{Schmidt2000} or even allow for the direct control of the spin state in semiconductors via external fields. The realization of such ``dilute magnetic semiconductors (DMS)'' is still a widely discussed issue in the multifunctional materials research community \cite{Dietl2010}. To this end, it is of particular importance to achieve room-temperature ferromagnetism and to avoid the formation of secondary magnetic phases. The latter are prone to significantly influence the magnetic properties of DMS \cite{Sato2007}, are difficult to detect \cite{Opel2008}, and lead to contradictory interpretations of the measured magnetic properties.

While ferromagnetism has first been observed in Mn-doped InAs \cite{Munekata1989} and GaAs \cite{Ohno1996} the Curie temperatures $T_\mathrm{C}$ in those dilute magnetic III-V systems do not exceed 170\,K \cite{Jungwirth2006}, making room temperature operation impossible. In contrast, ferromagnetism above 300\,K has been predicted for wide bandgap semiconductors such as ZnO or GaN substituted with magnetic ions \cite{Dietl2000}. This marked the starting point of an exciting race for room temperature DMS which pushed ZnO into the focus of materials research \cite{Ogale2010}. Early reports found indication for long-range magnetic order, even at room temperature, in ZnO doped with different $3d$ transition metals (TM) \cite{Pearton2004b}, including Sc \cite{Venkatesan2004}, Ti \cite{Venkatesan2004}, V \cite{Venkatesan2004,Saeki2001}, Mn \cite{Nielsen2005,Lim2004}, Fe \cite{Venkatesan2004}, Co \cite{Venkatesan2004,Ueda2001,Schwartz2004}, and Ni \cite{Venkatesan2004}. Since that time, there is an ongoing debate on whether these materials are really a DMS or whether the observed magnetic behavior is due to nanometer-sized precipitates of the magnetic TM dopant atoms embedded in the nonmagnetic ZnO matrix \cite{Coey2010}. During the past years, very detailed and thorough studies have provided increasing evidence that many of the previously published results on ferromagnetism in ZnO:TM could naturally also be explained in terms of such secondary TM phases \cite{Opel2008,Ney2010,Kaspar2008,Coey2008,Wi2004,Zhou2006,Jedrecy2009}, an effect which is also reported for other TM-substituted semiconductors like Ge:Mn \cite{Ahlers2006,Jager2006}. Some authors even discovered different magnetic regimes in ZnO:Co and ZnO:Mn, depending on the polarity \cite{Kittilstved2006a} or the density \cite{Behan2008} of the carriers. Others reported the possible control of ferromagnetism in ZnO:Co via the gate effect \cite{Lee2009} or by varying the grain boundary density \cite{Straumal2013}. In the following, we will summarize our broad research effort on Co, Cu, and (Li,Ni)-substitution in ZnO thin films.

\subsection{Cobalt Substitution.}
\label{sec:Co-doping}

ZnO:Co has been among the most intensely investigated DMS systems \cite{Ogale2010}. As pointed out above, the clarification of the origin of the observed magnetism together with the identification of the control parameters is difficult and still a matter of controversy \cite{Coey2008}. This is partly caused by the problem that different experimental groups have published incongruous results, ranging from room temperature ferromagnetism \cite{Kittilstved2006a,Pearton2004b,Venkatesan2004,Ueda2001,Schwartz2004,Gacic2008,Li2012,Chen2013b} to the absence of intrinsic DMS-type ferromagnetic interactions even at low temperatures \cite{Opel2008,Ney2010,Kaspar2008,Kolesnik2004,Lawes2005,Ney2008a,Tietze2008}. Applying element-specific synchrotron techniques such as X-ray magnetic circular dichroism (XMCD) led some authors to identify oxygen vacancies as the intrinsic origin for ferromagnetism \cite{Tietze2008} while others reported no ferromagnetism at all in their samples \cite{Opel2008,Ney2010,Kaspar2008}. To unambiguously clarify the nature of magnetism in ZnO:Co, we prepared thin films using laser-MBE \cite{Opel2012,Gross2000,Klein1999,Klein2000} and carefully investigated their structural and magnetic properties as discussed in brief in the following and in more detail in Ref.~\cite{Opel2008}.

\begin{figure}[tb]
    \includegraphics*[width=\linewidth]{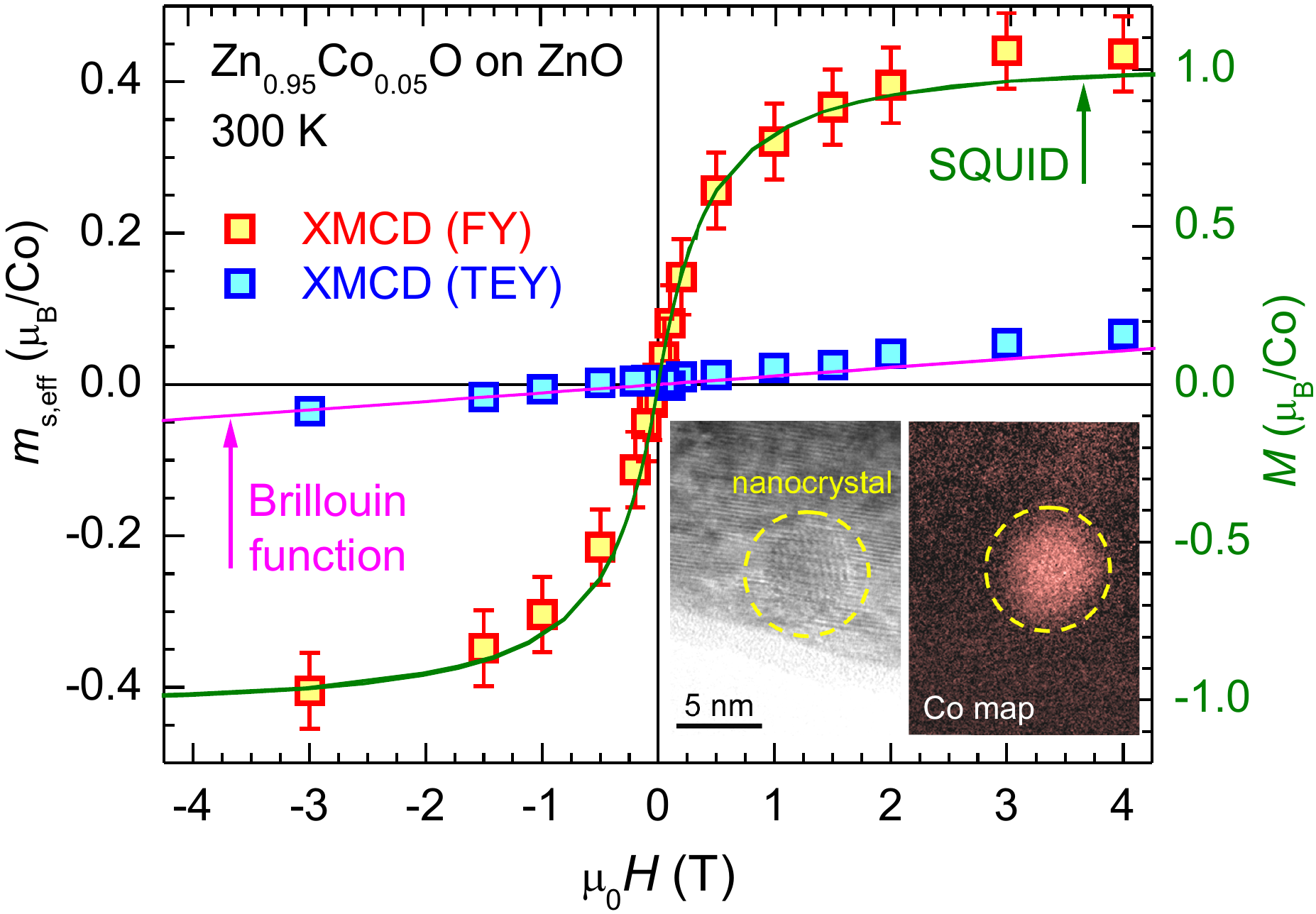}
    \caption[]{Effective spin magnetic moment $m_{\rm s,eff}$ of Co in
               Zn$_{0.95}$Co$_{0.05}$O, derived by X-ray magnetic circular dichroism (XMCD), recorded
               in total electron yield (TEY, blue) and fluorescence mode (FY, red). The FY
               data follow the sample magnetization $M$ (green), determined by SQUID
               magnetometry. The TEY data fit well to a Brillouin function for paramagnetic
               Co$^{2+}$ ions (pink). The inset shows a TEM image (left) displaying a
               nanocrystal in the ZnO matrix and an elemental map of Co obtained by EF-TEM
               (right), displaying Co enrichment in the same area. The figure is reproduced
               from Ref.~\cite{Opel2011}.}
    \label{fig:XMCD}
\end{figure}

Epitaxial, 350\,nm thin Zn$_{0.95}$Co$_{0.05}$O films were grown by pulsed laser deposition from a stoichiometric polycrystalline target using a KrF excimer laser (248~nm) at a repetition rate of 2~Hz with a laser fluence of 2~J/cm$^2$. The films were deposited \emph{homoepitaxially} on single crystalline, (0001)-oriented ZnO substrates at temperatures $T_\mathrm{sub}$ between $300^\circ$C and $600^\circ$C in Ar atmosphere at a pressure of $4 \times 10^{-3}$~mbar. Bulk ZnO crystallizes in the polar, hexagonal wurtzite structure (point group $C_{6v}$ or $6mm$, space group $C^4_{6v}$ or $P6_3mc$) with lattice parameters $a = 0.32501$\,nm and $c = 0.52071$\,nm \cite{Kisi1989}. We analyzed the Zn$_{0.95}$Co$_{0.05}$O film properties in a first comprehensive investigation \cite{Opel2008}, combining SQUID magnetometry, XMCD, and AC susceptibility measurements with careful X-ray and high resolution transmission electron microscopy (HR-TEM) studies. At room temperature, we find the highest magnetization of up to $1.95\,\mu_\mathrm{B}$/Co for thin films deposited at $T_\mathrm{sub} = 500^\circ$C. However, the $M(H)$ magnetization curves do neither display any evident ferromagnetic hysteresis, nor do they fit to a simple Brillouin function for distributed (diluted) Co$^{2+}$ paramagnetic moments (Fig.~\ref{fig:XMCD}). For an element-specific distinction between surface and bulk magnetic properties, we simultaneously recorded XMCD spectra in both the total electron (TEY) and the fluorescence yield (FY) modes, respectively (Fig.~\ref{fig:XMCD}). Our data provide clear evidence that our Zn$_{0.95}$Co$_{0.05}$O thin films are not homogeneous DMS. The large magnetic moments observed at room temperature can be traced back to metallic superparamagnetic Co nanocrystals with total magnetic moments of up to $5900\,\mu_\mathrm{B}$ and blocking temperatures up to 40\,K, embedded in the thin film. Further clear evidence for the presence of Co precipitates is provided by XMCD and direct imaging by energy-filtering TEM (EF-TEM) (Fig.~\ref{fig:XMCD}). Also, high-resolution X-ray diffraction (HR-XRD) diagrams display a low intensity, but finite reflection around $2\theta \simeq 44^\circ$ which can be assigned to various Co-rich compounds like fcc or hcp metallic Co, ZnCo$_2$O$_4$ or Co$_3$O$_4$ \cite{Opel2008}. Zn$_{0.95}$Co$_{0.05}$O thin films deposited \emph{heteroepitaxially} on (0001)-oriented Al$_2$O$_3$ or ScAlMgO$_4$ substrates are superparamagnetic only when grown at $T_\mathrm{sub} < 400^\circ$C. Otherwise, they show pure paramagnetism of isolated Co$^{2+}$ moments ($J=3/2$) from room temperature down to 5\,K \cite{Ney2010}. More details are published elsewhere \cite{Opel2008,Ney2010}.

To further clarify the issue of Co precipitates, a comprehensive set of ZnO:Co epitaxial thin film samples fabricated using three deposition methods in four different laboratories has been studied \cite{Ney2010}. All samples have been analysed applying an identical set of XMCD and SQUID measurement protocols. The key conclusion is that phase-pure, crystallographically excellent ZnO:Co is uniformly paramagnetic, irrespective of the preparation method. Ferromagnetic-like behavior is observed only for ZnO:Co samples showing extensive defect formation or phase separation. A theoretical analysis of this situation has been performed only recently \cite{Chakraborty2012}, although it was pointed out earlier that phase segregation might be a problem for DMS systems and result in the observation of extrinsic ``phantom magnetism'' at room temperature \cite{Coey2006}.

\subsection{Lithium and Nickel Codoping.}
\label{sec:Li-doping}

Following a recent report \cite{Kumar2010}, we also studied the possibility of (Li,Ni)-codoping to establish ferromagnetism together with $p$-type conductivity in ZnO. Again, we deposited thin films from stoichiometric targets with compositions Zn$_{0.98-x}$Li$_x$Ni$_{0.02}$O ($x=0, 0.02, 0.05, 0.09$) heteroepitaxially on (0001)-oriented Al$_2$O$_3$ substrates in O$_2$ atmosphere at $400^\circ$C under different growth conditions. Samples grown at low fluence (1\,J/cm$^2$), repetition rate (2\,Hz), and oxygen pressure ($10^{-3}$\,mbar) show excellent structural quality. For all $x$ values, the HR-XRD diagrams do not indicate any secondary phases nor metallic Ni precipitates. The (0002) and (0004) reflections from the Zn$_{0.98-x}$Li$_x$Ni$_{0.02}$O thin films are clearly visible with an intensity close to the (0006) and (00012) reflections from the Al$_2$O$_3$ substrate. The epitaxial films show a nearly perfect in-plane orientation as demonstrated by $\phi$-scans of the $(10\bar{1}1)$ reflections. Moreover, the films display a low mosaic spread as indicated by a narrow full width at half maximum (FWHM) of the $\omega$ rocking curves of the (0002) reflections down to $0.08^\circ$. Films grown at high fluence, repetition rate, and oxygen pressure (2.7\,J/cm$^2$, 10\,Hz, $5 \times 10^{-2}$\,mbar, i.e.~the same values as in Ref.~\cite{Kumar2010}), however, display a lower structural quality. Their FWHM of the $\omega$ rocking curve of the same reflection is by a factor of 4 larger, indicating a higher mosaic spread, and the intensities of the film reflections are by one order of magnitude below those from the first set of samples. The film with $x=2\%$ also shows indication for the presence of secondary phases.

\begin{figure}[tb]
    \includegraphics*[width=\linewidth]{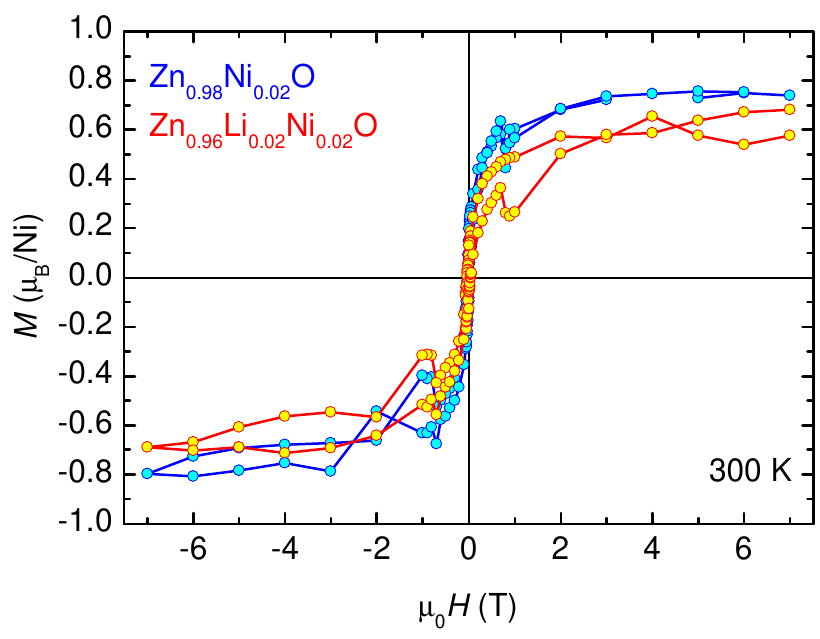}
    \caption[]{Room temperature SQUID magnetometry from thin films of Zn$_{0.98-x}$Li$_x$Ni$_{0.02}$O,
             deposited at high fluence/oxygen pressure, i.e.~the same values as in Ref.~\cite{Kumar2010}.
             For details see text.}
    \label{fig:senthil}
\end{figure}

The magnetization loops of the samples grown at high fluence/oxygen pressure show an ``S''-shaped behavior at room temperature (Fig.~\ref{fig:senthil}) which is reminiscent of our previous results for superparamagnetic Zn$_{0.95}$Co$_{0.05}$O described in the previous section (cf.~Fig.~\ref{fig:XMCD}) \cite{Opel2008,Ney2010}. Again, comparing $M(T)$ measurements after field cooling and zero-field cooling provides evidence for blocking behavior. The saturation magnetic moment at 7\,T does not increase significantly when cooling from 300\,K to 5\,K. Interestingly, its value is around $0.7\,\mu_\mathrm{B}$ per Ni and, hence, very close to the bulk value of Ni metal ($0.6\,\mu_\mathrm{B}$) but far from that of Ni$^{2+}$ ions ($2\,\mu_\mathrm{B}$). This suggests that again nanometer-sized metallic Ni precipitates are responsible for the observed room-temperature magnetism in these samples.

To clarify the issue of $p$-type conductivity and identify the sign of the charge carriers, we performed thermopower measurements to determine the Seebeck coefficient $S$ rather than Hall effect measurements. The latter are difficult to interpret if more than one conduction band is involved or if charge transport is via hopping. For all Zn$_{0.98-x}$Li$_x$Ni$_{0.02}$O thin film samples studied here, regardless of their different growth parameters, we find $S<0$ at room temperature indicating $n$-type conductivity. Within experimental error, $S$ ranges between $-400\,\mu{\rm V/K}$ and $-900\,\mu{\rm V/K}$, with smaller absolute values corresponding to lower Li concentrations. In summary, we cannot confirm $p$-type conductivity in any of our (Li,Ni)-substituted ZnO thin films.

\subsection{Copper Substitution.}
\label{sec:Cu-doping}

There exist several reports that ZnO turns ferromagnetic even after doping with \emph{nonmagnetic} elements, such as carbon \cite{Pan2007} or copper \cite{Tian2011}. For ZnO:Cu, the situation is controversial as ferromagnetism was found in macroscopic magnetization and magnetotransport measurements only, whereas element-specific, microscopic techniques confirmed a paramagnetic behavior of the Cu $3d$ states \cite{Keavney2007}. In general, when deriving ferromagnetic behavior from simple $M(H)$ magnetization curves, obtained with standard commercial SQUID magnetometer systems, particular care has to be taken before drawing conclusive results \cite{Stamenov2006}. Moreover, the tiny signals from thin film samples are superimposed by a huge diamagnetic or paramagnetic background from the underlying substrate with a much larger volume which has to be properly subtracted \cite{Ney2008c}. Possible magnetic impurities in the substrate represent an additional source for spurious magnetic signals. Here, we will demonstrate how $M(H)$ magnetization measurements can lead to the wrong conclusion of room temperature ferromagnetism in thin film samples of ZnO:Cu.

Epitaxial, 350\,nm thin Zn$_{0.95}$Cu$_{0.05}$O films were grown by pulsed laser deposition from a stoichiometric polycrystalline target using a KrF excimer laser (248~nm) at a repetition rate of 2~Hz with a laser fluence of 2~J/cm$^2$. The films were deposited heteroepitaxially on single crystalline, (0001)-oriented Al$_2$O$_3$ substrates at different temperatures $T_\mathrm{sub}$ between $300^\circ$C and $800^\circ$C in Ar atmosphere at a pressure of $4 \times 10^{-3}$~mbar. The HR-XRD diagrams demonstrate $c$-axis oriented, epitaxial growth of Zn$_{0.95}$Cu$_{0.05}$O. They show a weak reflection around $2\theta=34.5^\circ$ which can be assigned to Cu(111). Its intensity is three orders of magnitude smaller than that of ZnO(0002) for the sample grown at $T_\mathrm{sub} = 500^\circ$C and at least four orders of magnitude smaller for all other $T_\mathrm{sub}$. It indicates the presence of a secondary phase of metallic Cu, similar to our observations for ZnO:Co described in section~\ref{sec:Co-doping}.

\begin{figure}[tb]
    \includegraphics*[width=\linewidth]{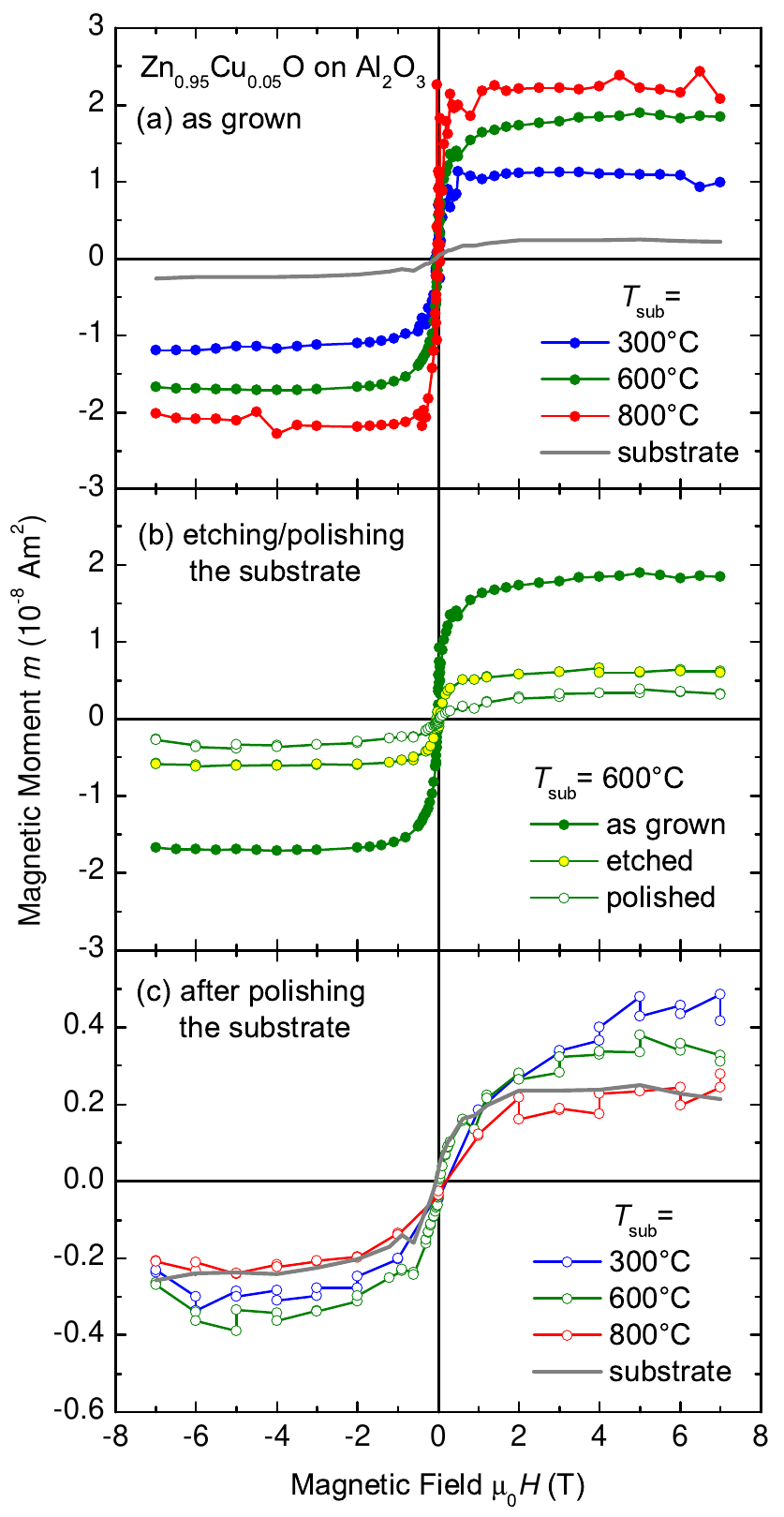}
    \caption[]{Room temperature SQUID magnetometry from thin films of Zn$_{0.95}$Cu$_{0.05}$O deposited at different substrate temperatures $T_\mathrm{sub}$. The diamagnetic contribution from the Al$_2$O$_3$ substrate has been subtracted from all curves.
               (a) Magnetization versus applied field directly after growth (symbols). The data from a bare substrate are shown for comparison (grey line).
               (b) Magnetization from the sample deposited at $600^\circ$C after growth, after chemical etching, and after mechanically polishing the back side of its substrate.
               (c) Magnetization of all samples from (a) after polishing the back side of the substrates.
               }
    \label{fig:CuSa}
\end{figure}

At room temperature, all films exhibit ferromagnetic-like behavior from SQUID magnetometry (Fig.~\ref{fig:CuSa}(a)). Their magnetization displays an ``S''-shape without magnetic hysteresis, inconsistent with a paramagnetic Brillouin function. Surprisingly, the saturation magnetic moment increases monotonically up to $2 \times 10^{-8}$\,Am$^2$ with increasing $T_\mathrm{sub}$. (We note that the bare Al$_2$O$_3$ substrate itself shows some magnetic saturation of $2 \times 10^{-9}$\,Am$^2$ as well.) This cannot be easily understood in a simple picture and points to an extrinsic origin of the observed magnetization. One possibility could be interdiffusion of magnetic ions from the metallic (steel) sample holder or the silver glue into the substrate from its back side at elevated deposition temperatures. To verify this assumption, we first chemically etched the back of the samples' substrates using HCl (38\%). By this step, the saturation magnetization significantly drops below $10^{-8}$\,Am$^2$, as is exemplarily shown for the sample deposited at $T_\mathrm{sub}=600^\circ$C in Fig.~\ref{fig:CuSa}(b). Then, we mechanically polished the back of the substrate, resulting in a further reduction of the saturation magnetization (Fig.~\ref{fig:CuSa}(b)). Applying this mechanical polishing procedure to the back sides of the substrates of all Zn$_{0.95}$Cu$_{0.05}$O samples, we are able to reduce their magnetic moments to the background level of the bare Al$_2$O$_3$ substrate of $2 \times 10^{-9}$\,Am$^2$ (Fig.~\ref{fig:CuSa}(c)). In conclusion, performing careful magnetic measurements and removing potential sources of systematic errors we cannot find any indication for any kind of magnetic signal from the Zn$_{0.95}$Cu$_{0.05}$O thin films. From these studies, we consider Zn$_{0.95}$Cu$_{0.05}$O nonmagnetic at room temperature.

\subsection{Conclusion.}

We cannot find any indication of ferromagnetism in Co-, (Li,Ni)-, and Cu-substituted ZnO. Our data do, of course, not prove that the realization of a ZnO-based DMS via doping with $3d$ transition metal ions is generally impossible. However, they clearly show that great care is required to unambiguously determine the nature of ferromagnetism in possible DMS systems and that reports on room-temperature ferromagnetism in TM doped ZnO should be taken with caution. In particular, conclusions based solely on $M(H)$ ``hysteresis'' curves are not reliable. Our ZnO:TM thin films so far behave either paramagnetic or superparamagnetic. None of the films show evidence for homogeneous ferromagnetic order as required for spintronics applications. We finally note that rare earth ion substitution in ZnO was considered to be an interesting alternative for achieving robust magnetic moments, but led to purely paramagnetic behavior as well \cite{Ney2012}.

\section{Spin Transport in ZnO.}
\label{sec:spintronics}

In conventional electronics, ZnO is widely used as a transparent conducting oxide (TCO) \cite{Grundmann2010}. It shows a direct and wide band gap of $E_g=(3.365 \pm 0.005)$\,eV at 300\,K in the near-ultraviolet range together with an electron mobility of 200\,cm$^2$/Vs and a large free-exciton binding energy of $(59.5 \pm 0.5)$\,meV \cite{Thomas1960,Mang1995,Srikant1998,Reynolds1999}. Furthermore, a high electron mobility of 180,000\,cm$^2$/Vs is reported in (Mg,Zn)O/ZnO heterostructures together with the observation of a fractional quantum Hall effect \cite{Tsukazaki2010}.  Although the realization of room-temperature ferromagnetism in ZnO is a hard problem, the small spin-orbit coupling \cite{Fu2008} in ZnO and correspondingly large spin coherence length \cite{Harmon2009} makes ZnO suitable for transporting or manipulating spins in thin film spintronic devices.

In the following, we first highlight the recent progress of ZnO thin film technology and device fabrication \cite{Opel2013}. Then, we will review our activities in the fields of injection, transport, and detection of spin-polarized ensembles of charge carriers in ZnO using an all-optical \cite{Kuhlen2013b,Kuhlen2013c}, an all-electrical \cite{Althammer2012}, or a combined electrical/optical injection/detection scheme.

\subsection{Fabrication of ZnO Thin Films and Heterostructures.}

Within the last decade there has been enormous progress in the growth of epitaxial ZnO thin films on Al$_2$O$_3$ substrates by laser-MBE and other techniques, in spite of the large lattice mismatch of $18.2\%$. In the following we only give a brief overview on the growth issues, details have been  published recently \cite{Opel2013}.

An important step in the growth of high quality ZnO thin film samples is the annealing of the Al$_2$O$_3$ substrate at $850^\circ$C in $10^{-3}$\,mbar O$_2$ for 1\,h. Then, we deposit ZnO in the same atmosphere at $T_\mathrm{sub} = 400^\circ$C, using a laser fluence of 1\,J/cm$^2$ at 2\,Hz. The quality of ZnO thin films can be further improved by the insertion of a ZnO or (Mg,Zn)O buffer layer between the film and the substrate. The structural characterization of our thin films shows that they grow with lattice parameters close to the bulk values. Laue oscillations in the HR-XRD diagrams and the narrow FWHM of the ZnO(0002) $\omega$ rocking curve of only $0.03^\circ$ indicate a coherent growth with low mosaic spread along the out-of-plane direction. For the FWHM of the ZnO($10\overline{1}1$) $\omega$ rocking curve, we achieve values of $0.27^\circ$ and $0.46^\circ$ for 1000\,nm and 120\,nm thin ZnO layers, respectively, which are comparable to recently published data \cite{Miyamoto2002,Kaidashev2003,Wassner2009}. Moreover, we find an in-plane epitaxial relationship of Al$_2$O$_3(0001)[11\overline{2}0] \|$ZnO$(0001)[10\overline{1}0]$. The analysis of the Hall effect suggests that in our unbuffered as well as in the buffered ZnO films, two different layers contribute to the Hall properties: a degenerate layer located at the interface between substrate and film and a semiconducting layer on top. The extracted residual carrier concentration of our $n$-type ZnO at room temperature is as low as $n = 4.5 \times 10^{16}$\,cm$^{-3}$ caused by unintentional Al doping, on par to other reported values \cite{Wenckstern2007}. In contrast, the Hall mobility in our samples is by one order of magnitude lower than the values reported by other groups. Temperature dependent photoluminescence experiments confirm the existence of donor-bound and free excitons. The donor-bound excitons are related to Al donors, verifying aluminum as the main impurity in our samples \cite{Kuhlen2013b}. Moreover, the FWHM of the donor-bound exciton line $I_6$ is as low as 3.5\,meV, equal to reported values \cite{Chen1998}.

For spintronic applications, charge carrier populations with controllable spin polarizations must be created in ZnO. A seemingly straightforward approach is to inject spin-polarized carriers from a ferromagnetic electrode into the semiconducting material. Unfortunately, the large conductance mismatch between conventional metallic $3d$ ferromagnets and semiconductors prevents an efficient spin injection \cite{Schmidt2000}. This problem can be circumvented by utilizing non-metallic ferromagnetic materials with low electrical conductivity. So-called ``half-metallic'' ferromagnets with a spin polarization of $100\%$ are thus most interesting. The oxide ferrimagnet Fe$_3$O$_4$ is such a half-metal according to band structure calculations \cite{Zhang1991}, and a spin polarization of up to $-(80 \pm 5)$\,\% has been reported from spin-resolved photoelectron spectroscopy in $(111)$-oriented Fe$_3$O$_4$ \cite{Fonin2005,Fonin2008}. Furthermore, we demonstrated that the electrical conductivity $\sigma \approx 200\,(\mathrm{\Omega cm})^{-1}$ of Fe$_3$O$_4$ at room temperature \cite{Reisinger2004,Venkateshvaran2008} is low and can be tuned in a wide range via Zn substitution \cite{Venkateshvaran2009} while its Curie temperature $T_{\rm C} \simeq 860$\,K is well above 300\,K, making Fe$_3$O$_4$ a promising material for spin injection into semiconductors. We also demonstrated that $(111)$-oriented Fe$_3$O$_4$ can be epitaxially grown onto ZnO using laser-MBE \cite{Boger2008}. We further demonstrated the epitaxial growth of ZnO thin films onto ferrimagnetic $(111)$-oriented Fe$_3$O$_4$ layers. A detailed characterization of the samples using HR-XRD, atomic force microscopy (AFM), TEM, and SQUID magnetometry is published elsewhere \cite{Boger2008} and shows that the magnetic and structural properties of our (111)-oriented Fe$_3$O$_4$ films on ZnO are state of the art with sharp Fe$_3$O$_4$/ZnO interfaces.

\subsection{Optical Injection -- Optical Detection.}
\label{sec:optical-optical}

The spin coherence time of mobile charge carriers -- and the associated length scale for coherent spin transport -- are fundamental parameters for semiconductor spintronics. They can be probed in an all-optical pump-probe scheme by time-resolved Faraday rotation (TRFR), where a spin polarization in the conduction band is created or detected via a pump or probe pulse with circular or linear polarization, respectively. By simultaneously applying a magnetic field, the precession of the spins between pump and probe can be observed via their Kerr ellipticity $\eta_\mathrm{K}$ \cite{Kuhlen2013c}. While the spin-coherent properties of e.g.~GaAs and related III-V semiconductor compounds were extensively studied in this way \cite{Awschalom2002}, only few reports exist for ZnO so far \cite{Ghosh2005,Liu2007,Jansen2008}. Electron spin coherence up to room temperature in epitaxial ZnO thin films was first observed by Ghosh \textit{et al.} with a spin coherence time of $T_2^* \approx 188$\,ps at room temperature and $T_2^* \approx 2$\,ns at 10\,K \cite{Ghosh2005}. In a similar TRFR experiment, we find longer $T_2^*$ times in our high-quality ZnO thin films \cite{Kuhlen2013b}. In addition to a precession signal with a long spin dephasing time of $T_2^* = 15$\,ns, we observe a shorter component with a $T_2^*$ of few nanoseconds. While the former can be attributed to electrons bound to Al donors, the latter is due to mobile charge carrier electrons as discussed below for the all-electrical spin injection experiments \cite{Althammer2012}. The polarization efficiency of each species shows different temperature and photon energy dependencies. While donor electrons can most efficiently be polarized at low temperatures and bound to neutral donors via exciton excitation and relaxation, the short component can be best observed via ionized donors and persists up to room temperature \cite{Kuhlen2013b}. Due to these dependencies we are able to control the polarization amplitude of each spin species in ZnO. More details are published elsewhere \cite{Kuhlen2013b,Kuhlen2013c,Kuhlen2013a}.

\subsection{Electrical Injection -- Electrical Detection.}
\label{sec:electrical-electrical}

Reports on electrical spin injection in ZnO, in general, are rare \cite{Chen2002,Ji2009,Shimazawa2010} and mainly focus on technical aspects rather than fundamental spin-dependent properties. To directly investigate the carrier spins in our ZnO thin films together with their spin relaxation, we performed electrical transport experiments in Ni/ZnO/Co spin valve multilayer structures \cite{Althammer2012}. The multilayer system was fabricated \textit{in-situ} on (0001)-oriented Al$_2$O$_3$ substrates and consists of a 12\,nm thin metallic TiN layer as bottom electrode, a 11\,nm thin Co layer as the first ferromagnetic electrode, a semiconducting ZnO layer with variable thickness from 15\,nm to 100\,nm, a 11\,nm thin Ni layer as the second ferromagnetic electrode, and finally a 24\,nm thin metallic Au layer as top electrode. TiN and ZnO were deposited by laser-MBE, Co, Ni, and Au by thermal evaporation, as described elsewhere \cite{Althammer2012,Opel2013}. Junctions with areas of $400\,\mu$m$^2$ were patterned by photolithography and investigated by magnetotransport.

\begin{figure}[tb]%
  \includegraphics*[width=\linewidth]{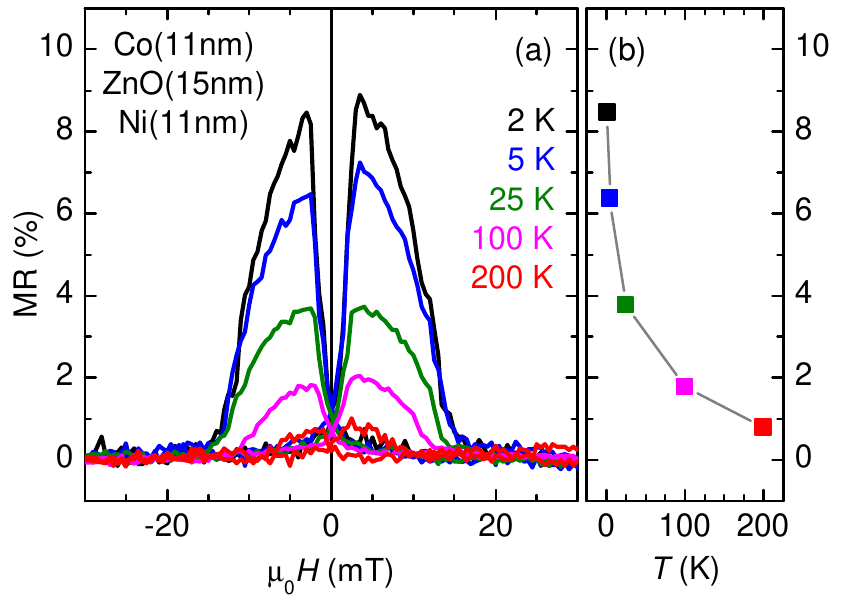}
  \caption[]{(a) Magnetoresistance MR of a Co/ZnO/Ni spin valve structure versus magnetic field $H$ applied in-plane for different temperatures $T$.
             (b) Temperature dependence of the maximum MR.}
    \label{fig:electrical-electrical}
\end{figure}

The magnetoresistance (MR) displays two resistive states (Fig.~\ref{fig:electrical-electrical}) with switching fields corresponding to the coercivities of the respective ferromagnets Co and Ni \cite{Althammer2012}. This behavior is evident for a spin valve effect and demonstrates the successful electrical creation and detection of a spin-polarized ensemble of electrons in ZnO, including the transport of this spin information across several nanometers in ZnO. We note that a tunneling magnetoresistive (TMR) effect can be safely excluded due to the large thicknesses of the ZnO layers of $t_\mathrm{ZnO} \geq 15$\,nm. We obtain the maximum MR of $8.4\%$ for $T = 2$\,K and $t_\mathrm{ZnO} = 15$\,nm. With increasing temperature, our MR decreases to $1.8\%$ at 100\,K and $0.8\%$ at 200\,K (cf. Fig.~\ref{fig:electrical-electrical}). Compared to data reported in the literature ($1.38\%$ and $1.12\%$ at 90\,K for $t_\mathrm{ZnO} = 3$\,nm and 10\,nm \cite{Ji2009}), our MR values are by a factor of two higher. Analyzing our data from samples with different $t_\mathrm{ZnO}$ in a Valet-Fert approach \cite{Valet1993,Fert2001}, we obtain spin diffusion lengths of 10.8\,nm (2\,K), 10.7\,nm (10\,K), and 6.2\,nm (200\,K) as well as spin lifetimes of 2.6\,ns (2\,K), 2.0\,ns (10\,K), and 31\,ps (200\,K) \cite{Althammer2012}, corresponding to the short component from the above described TRFR analysis \cite{Kuhlen2013b}. More details are published elsewhere \cite{Althammer2012}.

\subsection{Electrical Injection -- Optical Detection.}
\label{sec:electrical-optical}

\begin{figure*}[tb]%
  \sidecaption
  \includegraphics*[width=.67\textwidth]{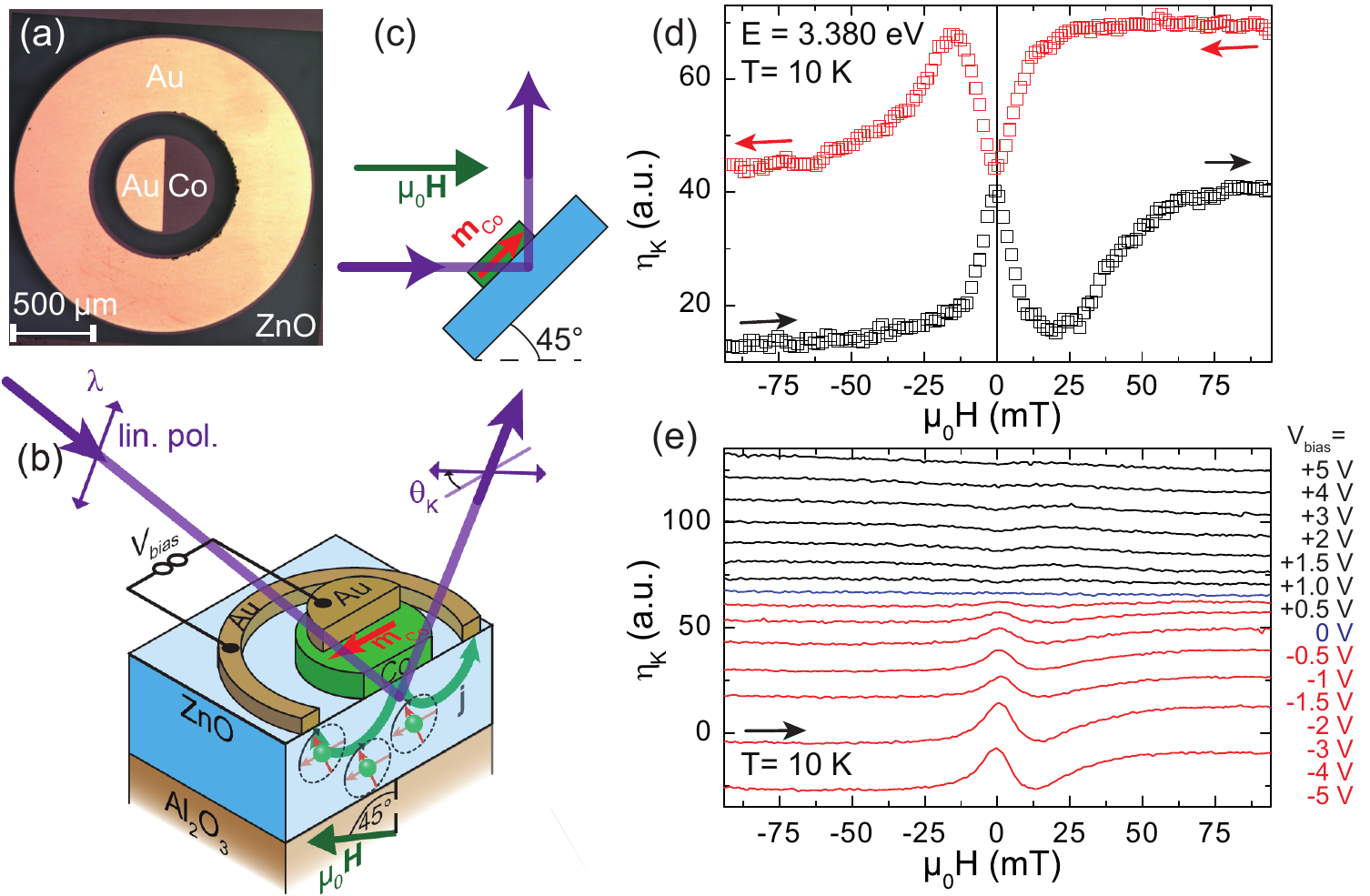}
  \caption[]{(a) Optical micrograph of the mesa structure for electrical injection/optical detection experiments.
             (b) Illustration of the measurement setup. At reverse bias a spin-polarized current is injected into ZnO. The spin polarization is detected optically by the Kerr ellipticity $\eta_\mathrm{K}$ of the reflected linearly polarized light.
             (c) Orientation of magnetization $\vec{m}_\mathrm{Co}$ and external field $\mu_0\vec{H}$ in an oblique configuration.
             (d) $\eta_\mathrm{K}$ for the up (black) and down sweep (red) at 10\,K and $V_\mathrm{bias} = -5$\,V.
             (e) $\eta_\mathrm{K}$ vs.~$\mu_0H$ for different $V_\mathrm{bias}$ at 10\,K.
           }
  \label{fig:electrical-optical-bias}
\end{figure*}

In the following, we present experiments on electrical injection and optical detection which allow us to determine electron spin dephasing times in ZnO after electrical spin injection. The sample consists of a 30\,nm thin Co film which was grown \textit{in-situ} by electron beam evaporation on a 160\,nm thick ZnO layer. We first defined a disc-shaped Co electrode with a diameter of 400\,$\mu$m (cf. Fig.~\ref{fig:electrical-optical-bias}(a)). In a second photolithography step, we patterned a semicircular Au contact on top together with a ring-shaped outer Au contact by DC-sputtering and lift-off (Fig.~\ref{fig:electrical-optical-bias}(a,b)).

The magneto-optical detection scheme is illustrated in Fig.~\ref{fig:electrical-optical-bias}(b). By applying a dc voltage $V_\mathrm{bias}$ between the outer Au contact and the inner Au/Co electrode, we inject electron spins from Co into ZnO (green arrows). Their initial spin orientation near the Co/ZnO interface is defined by the magnetization direction of the ferromagnet. Subsequent spin precession in an external magnetic field can only be observed if the magnetic field direction is non-collinear to the spin orientation. In dc experiments, spin precession results in a rapid depolarization of the steady-state spin polarization (the Hanle effect \cite{Hanle1924}), because spins are being continuously injected in the time-domain.

As the Co layer is polycrystalline, it exhibits no in-plane magnetic anisotropy. For spin precession, we thus used the out-of-plane magnetic anisotropy by tilting the sample by $45^\circ$ with respect to both magnetic field and laser probe direction (Fig.~\ref{fig:electrical-optical-bias}(c)). The latter is linearly polarized, focused on the Co electrode, and stems from a frequency doubled Ti:sapphire laser with tunable laser energy. The reflected laser beam was analyzed for its Kerr ellipticity $\eta_\mathrm{K}$, which is a direct measure of the spin orientation along the incident probe laser beam direction.

At moderate magnetic fields, electron spins will point almost parallel to the sample plane after injection and, thereafter, can precess about the external magnetic field. For our device, we observe a clear symmetric Hanle depolarization curve in the Kerr ellipticity $\eta_\mathrm{K}$ for $|\mu_0H| < 10$\,mT, as shown in Fig.~\ref{fig:electrical-optical-bias}(d) for $E=3.38$\,eV at 10\,K. We note that the background ($|\mu_0H| > 10$\,mT) is hysteretic and results from the longitudinal spin component, i.e. the one parallel to the magnetic field direction which switches during magnetization reversal. As expected, the Hanle curve inverts after changing the sweep direction at external fields larger than the coercivity of the Co electrode, as the magnetization direction and therefore the spin polarization of the Co injector is inverted. While spins are injected from Co into ZnO at reverse bias ($V_\mathrm{bias}<0$\,V), they are extracted from ZnO into Co for positive bias. For the latter case, the remaining spin polarization in the ZnO layer has reversed spin orientation which is evident by the observed sign reversal in Fig.~\ref{fig:electrical-optical-bias}(e). However, the Hanle signals for $V_\mathrm{bias}>0$ are weak as the spin extraction from ZnO into Co is inefficient. At $V_\mathrm{bias}=0$, no Hanle signature is detected (blue curve in Fig.~\ref{fig:electrical-optical-bias}(e))

\begin{figure}[b]%
  \includegraphics*[width=\linewidth]{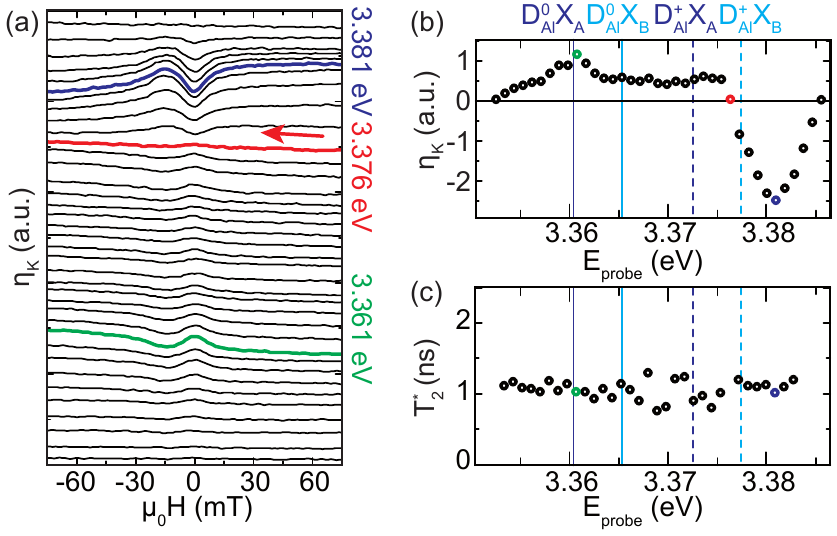}
  \caption[]{(a) Kerr ellipticity $\eta_\mathrm{K}$ vs.~external magnetic field $\mu_0H$ of a Co/ZnO spin injection device for different laser probe energies $E_\mathrm{probe}$ at one magnetic field sweep direction. All spectra are displayed by a vertical offset for clarity.
             (b) Amplitude of Hanle peak vs. laser probe energy. The energies for both neutral and ionized donor bound excitons which are excited from both the A and B valence band of ZnO at Al doping sites are included as a reference \cite{Kuhlen2013b}.
             (c) The spin dephasing times $T_2^*$ are independent of the laser probe energies.
           }
  \label{fig:electrical-optical-hanle}
\end{figure}

The widths of the Hanle curves determine the spin dephasing times $T_2^*$ which we estimate to be 1\,ns. This value is independent of the laser detection energy as seen in Fig.~\ref{fig:electrical-optical-hanle}(c), where we cover the whole energy range of neutral and ionized donor bound exciton states as measured in TRFR discussed in section~\ref{sec:optical-optical} \cite{Kuhlen2013b}. In Fig.~\ref{fig:electrical-optical-hanle}(a), we show the corresponding Hanle curves for one sweep direction. The measured spin dephasing times are identified as the short component in TRFR, which demonstrates that the respective electronic states are mobile carriers rather than bound states. We furthermore notice that the measured Kerr ellipticity peaks (Fig.~\ref{fig:electrical-optical-hanle}(b)) are located at identical energies as in TRFR experiments \cite{Kuhlen2013b}.

\section{Summary.}

The extensive research on zinc oxide in the past years has discovered many new fascinating aspects which are important for its possible use in future spintronic applications. On the one hand, dilute magnetic doping of ZnO, initially considered a success story, turned out to be more difficult than expected. First positive reports have to be re-considered in the light of element-specific, microscopic techniques applied to such ``ferromagnetic'' transition metal-doped oxide materials. The latter experiments reveal that phase separation and defect formation are responsible for the observed macroscopic magnetic signals in most ZnO-based compounds. Our ZnO:Co, ZnO:Cu, and ZnO:(Li,Ni) thin film samples are either paramagnetic or superparamagnetic due to nanometer-sized metallic inclusions. None of them displays any signature for intrinsic ferromagnetism at room temperature. On the other hand, spin coherence and spin dephasing properties of ZnO were studied applying both optical and electrical spin injection and detection schemes. From our detailed experimental investigation, we find a long ($\sim\,15$\,ns) and a short component ($\sim\,2$\,ns) of the spin dephasing time in our ZnO thin films at low temperatures. The long component is attributed to electrons localized at Al donors and is only seen in optical pump/optical probe experiments. The short one is observed in an ``electrical pump''/optical probe investigation as well as in all-electrical magnetotransport experiments, and reflects the behavior of mobile charge carriers.

\begin{acknowledgement}
We thank T.~Brenninger and S. Gepr\"{a}gs for valuable technical support and A.~Erb for the preparation of the polycrystalline PLD targets. This work was supported by the Deutsche Forschungsgemeinschaft (DFG) via SPP~1285 (projects No.~GR~1132/14 and BE~2441/4), the German Excellence Initiative via the ``Nanosystems Initiative Munich (NIM)'', and the European Synchrotron Radiation Facility (ESRF) via HE-2089.
\end{acknowledgement}

%
\bibliographystyle{pss}
\bibliography{SPP1285-Gross-Goennenwein-Opel}
%

%
%

\end{document}